\documentclass[nofootinbib,notitlepage,superscriptaddress,10pt,aps,pra,twocolumn]{revtex4-1}

\usepackage[utf8x]{inputenc}
\usepackage{amsmath}
\usepackage{amssymb}
\usepackage{graphicx}

\begin{document}

\title{On the UV dimensions of Loop Quantum Gravity}

\author{Michele RONCO}
\affiliation{Dipartimento di Fisica, Universit\`a di Roma ``La Sapienza", P.le A. Moro 2, 00185 Roma, Italy}
\affiliation{INFN, Sez.~Roma1, P.le A. Moro 2, 00185 Roma, Italy}

\begin{abstract}
\noindent Planck-scale dynamical dimensional reduction is attracting more and more interest in the quantum-gravity literature since it seems to be a model independent effect. However different studies base their results on different  concepts of spacetime dimensionality. Most of them rely on the \textit{spectral} dimension, others refer to the \textit{Hausdorff} dimension and, very recently, it has been introduced also the \textit{thermal} dimension. We here show that all these distinct definitions of dimension give the same outcome in the case of the effective regime of Loop Quantum Gravity (LQG). This is achieved by deriving a modified dispersion relation from the hypersurface-deformation algebra with quantum corrections. Moreover we also observe that the number of UV dimensions can be used to constrain the ambiguities in the choice of these LQG-based modifications of the Dirac spacetime algebra. In this regard, introducing the \textit{polymerization} of connections i.e. $K \rightarrow \frac{\sin(\delta K)}{\delta}$, we find that the leading quantum correction gives $d_{UV} = 2.5$. This result may indicate that the running to the expected value of two dimensions is ongoing, but it has not been completed yet. Finding $d_{UV}$ at ultra-short distances would require to go beyond the effective approach we here present. 
\end{abstract}

\maketitle

\section{INTRODUCTION}

There is an increasing interest in the quantum-gravity literature about the effect of dynamical dimensional reduction   of spacetime. It consists in a scale dependence of the dimension $d$ that runs from the standard IR value of four spacetime dimensions to the lower value $d \simeq 2$ at Planckian energies. Remarkably, despite the fact that quantum-gravity approaches start from different conceptual premises and adopt different formalisms, this dimensional running has been found in the majority of them, such as Causal Dynamical Triangulations (CDT) ~\cite{cdt}, Horava-Lifshitz gravity ~\cite{horava}, Causal Sets ~\cite{sets}, Asymptotic Safety ~\cite{asysaf}, Spacetime Noncommutativity ~\cite{ncst} and LQG ~\cite{lqg,lqg1,calcoriti}, which is here of interest. \\
However, in quantum gravity even the concept of spacetime dimension is a troublesome issue and it requires some carefulness. In fact, non-perturbative, background independent approaches (e.g. LQG ~\cite{lqg2,lqg3} and CDT ~\cite{cdt1}) generally rely on non-geometric quantities and they have discreteness as their core feature. For this reason, in order to extract phenomenological predictions, it would be necessary a coarse-graining process aimed at deriving a more manageable effective description from the fundamental discrete blocks, that characterize the Planckian realm. It is a common expectation that this procedure should leave some traces in a semiclassical regime where the emerging picture would be given in terms of a quantum spacetime. This reduction has the advantage of allowing us to recover at least some of our more familiar physical observables or, when it would not be possible, analogous ones with potential departures from their classical counterparts. The dimension belongs to this latter set of semiclassical observables because the usual Hausdorff dimension is ill-defined for a quantum spacetime ~\cite{carlip}. In the CDT approach ~\cite{cdt,cdt1} it was recognized for the first time that a proper "quantum analogue" could be the spectral dimension $d_{S}$, which is the scaling of the heat-kernel trace and it reproduces the standard Hausdorff dimension when the classical smooth spacetime is recovered. What is more, it was found that in the UV $d_S \simeq 2$ (see however Ref. ~\cite{cdt2} for recent CDT simulations favouring a smaller value of the dimension), which is now a recurring number in the literature ~\cite{dim1,dim2,dim3,dim4}. In the asymptotic safety program such a value is also intimately connected to the hope of having a fixed point in the UV. In fact, it has been proven that renormalizability is accomplished only if the dimension runs to two ~\cite{Percacci}. Furthermore, this prediction finds support in a recently developed approach ~\cite{padma} that has the advantage of relying on a minimal set of assumptions. Provided that quantum gravity will host an effective limit characterized by the presence of a minimum allowed length (identified with the Planck length), then it is shown ~\cite{padma} that the Euclidean volume becomes two-dimensional near the Planck scale \footnote{The author is grateful to Thanu Padmanabhan for pointing this out.}.  \\However, in a recent paper ~\cite{termdim} the physical significance of $d_S$ has been questioned. Such a concern is based on two observations: the computation of $d_S$ requires a preliminary Euclideanization of the spacetime and also it turns out to be invariant under diffeomorphisms on momentum space. Both these features are regarded as an evidence of the fact that $d_S$ is an unphysical quantity ~\cite{termdim}. Given that, it has been proposed to describe the phenomenon of dimensional reduction in terms of the thermal (or thermodynamical) dimension $d_T$, which can be defined as the exponent of the Stefan-Boltzmann law. Then, the UV flowing of $d_T$ is realized through a modified dispersion relation (MDR) that affects the partition function used to compute the energy density (see Ref. ~\cite{termdim} or Section III for further details). Thus, the value of $d_T$ near the Planck scale depends crucially on the specific form of the MDR. Furthermore, it has been recently noticed (see Refs. ~\cite{momhaus,momhaus1})  that form the MDR is  also possible to infer the Hausdorff dimension $d_H$ of energy-momentum space. If the duality between spacetime and momentum space is preserved in quantum gravity, this framework should provide another alternative characterization of the UV running. 
In this way we are in presence of a proliferation of distinct descriptions of the UV dimensionality of a quantum spacetime. These pictures make use of very different definitions of the dimension and, in principle, there is no reason why they should give the same outcome. On the other hand, they all coincide in the IR-low-energy regime where they reduce to $4$ and, thus, we could expect that this should happen also in the UV.\\ 
In this paper we show that this advisable convergence can be achieved in the semi-classical limit of LQG under rather general assumptions. The insight we gain is based on the recently proposed quantum modifications of the hypersurface deformation algebra (or the algebra of smeared constraints) ~\cite{hyp1,hyp2,hyp3,hyp4,hyp5}, which reduces to a correspondingly deformed Poincaré algebra in the asymptotic region, as shown in Ref. ~\cite{hyppa}. These Planckian deformations of the relativistic symmetries  are a key feature of the Deformed Special Relativity scenario ~\cite{dsr,dsr1}, as already pointed out in Ref. ~\cite{hyppa}, and, what is more, it has been recently shown in Ref. ~\cite{kappalqg} that they are consistent with a $\kappa$-Minkowski noncommutativity of the spacetime coordinates ~\cite{majRue,lukRueg}. We here exploit them to compute the MDR, thereby linking the LQG-based quantum corrections to the deformation of the dispersion relation. The general form of the MDR we derive allows us to find that both the spectral, the thermal and the Hausdorff\footnote{Notice that we are always referring to the Hausdorff dimension of momentum space since, as we mentioned, that of a quantum spacetime cannot be defined.} dimensions follow the same UV flowing, i.e. $d_S = d_T = d_H$.\\
Another significant observation we make is that, following our analysis "in reverse", we can get information about the LQG quantum-geometric deformations, that affect the Dirac algebra, from the value of the UV dimension. The importance of this recognition resides on the fact that these modifications are subjected to many sources of possible ambiguities ~\cite{amb,amb1,amb2} coming e.g. from the regularization techniques used to formally quantize the Hamiltonian constraint. These ambiguities are far from being resolved and it is still no clear if they may affect potentially physical outcomes ~\cite{amb3}. Thus, they are usually addressed only on the basis of mere theoretical arguments or being guided by a principle of technical simplification. The main sources of ambiguities are the spin representations of the quantum states of geometry as well as the choice of the space lattice and, in effective models we will here consider, they correspond respectively to holonomy and inverse-triad corrections. We here partially fix them with what is believed to be a phenomenological prediction i.e. the number of UV dimensions. Notably, we notice that the leading order correction provided by the holonomy corrections of homogeneous connections ~\cite{holo,holo1}, which are often implemented by simply taking the expression $\frac{\sin(\delta K)}{\delta}$ instead of $K$, e.g. in Loop Quantum Cosmology (LQC) ~\cite{lqc,lqc1} (see also ~\cite{aba} for a recent review on symmetry reduced models of LQG), is compatible with $d_{UV} = 2.5$. Thus, as we would have expected, the number of dimensions is correctly flowing to lower values even if it has not already reached the value of $2$, a value which is favoured in the quantum-gravity literature for the aforementioned reasons. On the basis of the steps we sketch out in the analysis we are here reporting it should be possible to exclude all the deformation functions $f(K)$ that are not consistent with $d_{UV} = 2$.  Remarkably, those quantum corrections, which are related to LQC as well as to the semi-classical limit of the theory, seem to point toward the right UV flowing. As already stated, the prediction of Planckian dimensional reduction has been also confirmed by previous LQG analyses ~\cite{lqg,lqg1} but more refined computations of Ref. ~\cite{calcoriti} have revealed that the "magic" number of $2$ can be reproduced only focusing on a specific superposition of kinematical spin-networks states (see ~\cite{calcoriti} for the details). Relying on the recently developed effective methods for LQG, we here provide further support to the idea that the effective spacetime of LQG maybe two dimensional at ultra-Planckian scales.

\section{MODIFIED DISPERSION RELATION}

We start considering the classical hypersurface deformation algebra (HDA), which was first introduced by Dirac ~\cite{dirac}. It is the set of Poisson brackets closed by the smeared constraints of the Arnowitt-Deser-Misner formulation of General Relativity (GR) ~\cite{adm}. The Dirac algebra of constraints is the way in which general covariance is implemented once the spacetime manifold has been split into the time direction and the spatial three surfaces i.e. $\mathcal{M} = \mathbb{R}\times \Sigma$. \\It is given by:
\begin{equation}
\begin{split}
\{D[M^{k}],D[N^{j}]\}=D[\mathcal{L}_{\vec{M}}N^{k}],\\
\{D[N^{k}],H[M]\}=H[\mathcal{L}_{\vec{N}}M],\\ 
\{H[N],H[M]\}=D[h^{jk}(N\partial_{j} M-M\partial_{j} N)],
\end{split}
\end{equation}
where $H[N]$ is the Hamiltonian (or scalar) constraint, while $D[N^{k}]$ is the momentum (or vector) constraint. The function $N$ is called the lapse and it is needed to implement time diffeomorphisms, while $N^{i}$ is the shift vector necessary to move along a given hyper-surface and, finally, $h^{ij}$ is the inverse metric induced on $\Sigma$. Thus, $H[N]$ and $D[N^{k}]$ have to be understood as the generators of gauge transformations which, in the case of GR, are space-time diffeomorphisms. 
For the purposes of our analysis, it is relevant the well established fact (see Ref. ~\cite{reggeteit}) that, when the spatial metric is flat $h_{ij} = \delta_{ij}$, if we take $N = \bigtriangleup t+v_{a}x^{a}$ (where $v_a$ is the infinitesimal boost parameter) and $N^{k} = \bigtriangleup x^{k}+R^{k}_{l}x^{l}$ (where $R^{k}_{l}$ is the matrix that generates infinitesimal rotations), we can infer the Poincaré algebra from the Dirac algebra ~\cite{reggeteit}. This classical relation is expected to hold also at the quantum level.\\
One of the open issues in the LQG research is the search for fully quantized versions of the constraints $H[N], D[N^{k}]$ on a Hilbert space. While it is known how to treat spatial diffeomorphisms and also how to solve the momentum constraint ~\cite{diff,diff1} thereby obtaining the kinematical Hilbert space of the theory, the finding of a Hamiltonian operator is far from being completed. However, over the last fifteen years several techniques have been developed, using both effective methods and discrete operator computations. In this way some candidates for an effective scalar constraint $H^{Q}[N]$ have been identified ~\cite{ham1,ham2,ham3}. For the analysis we are here reporting, the interesting fact is that the semi-classical corrections introduced in the Hamiltonian leave trace in the algebra of constraints . Remarkably, even if these calculations use different formalisms and they are based on different assumptions, the general form of the modified HDA turns out to be the same in all these studies, i.e. only the Poisson bracket between two scalar constraints is affected by quantum effects ~\cite{holo1,ham4}: 
\begin{equation}
\label{defhda}
\{H^{Q}[M],H^{Q}[N]\}=D[\beta h^{ij}(M\partial_{j}N-N\partial_{j}M)]
\end{equation}
where the specific form of the deformation function $\beta$ as well as its dependence on the phase space variables, which are $(h_{ij},\pi_{ij})$ if we use the metric formulation or $(A^{a}_{i}, E^{i}_{a})$ if we use the Ashtekar's one, varies with the quantum corrections considered to define $H^{Q}[N]$.\\
One of the causes of these quantum modifications of the scalar constraint is the fact that LQG cannot be quantized directly in terms of the Ashtekar variables $A^{a}_{i}$, which have to be replaced with their parallel propagators (or holonomies)~\cite{lqg2,lqg3,holo,holo1} $h_{\alpha}(A) = \mathcal{P}e^{-\int_{\alpha}t^{i}A^{a}_{i}\tau_{a}}$ (where $\mathcal{P}$ is the path-ordering operator, $t^{i}$ the tangent vector to the curve $\alpha$ and $\tau_a = -\frac{i}{2}\sigma_a$ the generators of $SU(2)$). If $a$ is the spatial index of a direction along which spacetime is homogeneous, then one has to consider just the local point-wise holonomies $h_{i}(A) = \cos(\frac{\delta A}{2})\mathbb{I}+\sin( \frac{\delta A}{2})\sigma_i$ (where $\delta \propto l_P = \sqrt{G} \approx 10^{-35} m$ is connected to the square root of the minimum eigenvalue of the area operator ~\cite{lqc1}).
These are the kinds of quantum effects considered in effective (semi-classical) LQG theories as well as both in spherically reduced models and in cosmological contexts ~\cite{aba}. In particular, for spherically symmetric LQG (see e.g. Refs. ~\cite{hyppa, ham3, ham4}) the deformation function depends on the homogeneous angular connection $K_{\phi}$ and it is directly related to the second derivative of the square of the holonomy correction $f(K_{\phi})$, i.e. $\beta = \frac{1}{2}\frac{d^{2}f^{2}(K_{\phi})}{dK^{2}_{\phi}}$.
Then, the important contribution of Ref. ~\cite{hyppa} has been to establish a link between these LQG-inspired quantum corrections and DSR-like deformations of the relativistic symmetries, thanks to the recognition that the angular connection $K_\phi \propto P_r$ is proportional to the Brown-York radial momentum ~\cite{BY} that generates spatial translations at infinity (see ~\cite{hyppa,kappalqg} for the details).
In fact, it has been shown that, taking the Minkowski limit of Eq. \eqref{defhda} as we sketched above for the classical case, the LQG-deformed HDA produces a corresponding Planckian deformation of the Poincaré algebra: 
\begin{equation}
\label{defpa}
[B_{r}, P_{0}] = iP_{r}\beta(l_{P} P_{r}) 
\end{equation}
where the other commutation relations remain unmodified. Here $B_r$ is the generator of radial boosts and $P_0$ the energy. The explicit form of $\beta$ is unknown and, as we already stressed, it is affected by ambiguities. In light of this, for our analysis we assume a rather general form:
\begin{equation}
\label{beta}
\beta(\lambda P_{r}) = 1+\alpha l_{P}^{\gamma} P^{\gamma}_{r} 
\end{equation} 
which is motivated by the above considerations and, obviously, satisfy the necessary requirement: $\lim_{l_P\to 0}\beta(l_P P_r) = 1$, i.e. we want to recover the standard Poincaré algebra in the continuum limit. We leave unspecified the constants of order one $\alpha$ and $\gamma$ that parametrize the aforementioned ambiguities. These parameters should encode at least the leading-order quantum correction to the Poincaré algebra.  \\
Using Eqs. \eqref{defpa}\eqref{beta} and taking into account that $[B_{r}, P_{r}] = i P_0, [P_{r}, P_{0}] = 0$, a straightforward computation gives us the following MDR: 
\begin{equation}
\label{mdr}
E^{2} = p^{2}+\frac{2\alpha}{\gamma +2}l_{P}^{\gamma} p^{\gamma+2}
\end{equation}
This completes the analysis started in Ref. ~\cite{hyppa} and carried on in Ref. ~\cite{kappalqg}, that aimed at building a bridge between the formal structures of loop quantization to the more manageable DSR scenario with the objective to enhance to possibilities to link mathematical constructions to observable quantities. The remarkable fact of having derived Eq. \eqref{mdr} form the LQG-deformed algebra of constraints \eqref{defhda} is that it will give us the opportunity to constrain experimentally the formal ambiguities of the LQG approach exploiting the ever-increasing phenomenological implications of MDRs (see Ref. ~\cite{gacLRR} and references therein). We are often in the situation in which quantum-gravity phenomenology misses a clear derivation from full-fledged developed approaches to quantum gravity or, on the contrary, the high complexity of these formalisms does not allow to infer testable effects. Following the work initiated in Ref. ~\cite{kappalqg}, we are here giving a further contribution to fill this gap.  
We find also interesting to notice that our MDR confirms a property of two previously proposed MDRs (see Refs.  ~\cite{GambiniPullin,AlfaroTecotl}), i.e. LQG corrections affect only the momentum sector of the dispersion relation leaving untouched the energy dependence. 
Therefore, this property, that has a rigorous justification in the spherically symmetric framework ~\cite{hyppa, ham3, ham4} we are here adopting, seems to be a recurring feature of LQG. Moreover, all the precedent analyses were confined to the kinematical Hilbert space of LQG, while we have here obtained Eq. \eqref{mdr} from the flat-spacetime limit of the full HDA including also the semi-classical Hamiltonian constraint \eqref{defhda}. Thus, even if we are working \textit{off-shell} (i.e. we do not solve the constraint equations), the MDR \eqref{mdr} should contain at least part of the dynamical content of LQG.  In the next section, we shall see that the form of Eq. \eqref{mdr} is crucial to prove that the running of dimensions does not depend on the chosen definition of the dimension.

\section{DIMENSIONS AND QUANTUM CORRECTIONS}

Our next task is to use the MDR \eqref{mdr} we derived in Section II in order to show that, regardless of the value of  the unknown parameters $\alpha$ and $\gamma$, the different characterizations of the UV flowing introduced in the literature predict the same number of dimensions if we consider the effective regime of LQG in the sense introduced in Refs. ~\cite{hyppa, ham3, ham4} and sketched in the previous section. 
To see this we start by the computation of the spectral dimension, which is defined as follows: 
\begin{equation}
d_S = -2\lim_{s\to 0}\frac{d \log P(s)}{d \log(s)}
\end{equation}
where $P(s)$ is the average return probability of a diffusion process in a Euclidean spacetime with fictitious time $s$. Following Refs. ~\cite{termdim,modesto,calcmod,spec1,spec2,calca}, we compute $d_S$ from the Euclidean version of our MDR \eqref{mdr} which is a d'Alembertian operator on momentum space:
\begin{equation}
\bigtriangleup^E = E^2+p^2+\frac{2\alpha}{\gamma +2}l_{P}^{\gamma} p^{\gamma+2}
\end{equation}
Then, a lengthy but straightforward computation (see ~\cite{spec1,spec2,calca}) leads to the following result: 
\begin{equation}
\label{spectral}
d_S = 1 + \frac{6}{2+\gamma}
\end{equation}
Notice that the value of $d_S$ does not depend on $\alpha$ but only on $\gamma$, i.e. only on the order of Planckian correction to the dispersion relation (see Eq. \eqref{mdr}). We will use this fact later on. \\
Now we want to show that also the thermal dimension $d_T$ is given by Eq. \eqref{spectral}. To this end we remind the definition of $d_T$ introduced in Ref.  ~\cite{termdim}. If you have a deformed Lorentzian d'Alembertian $\bigtriangleup^L_{\gamma_{t}\gamma_{x}} = E^2-p^2+l^{2\gamma_t}_t E^{2(1+\gamma_t)}-l^{2\gamma_x}_x p^{2(1+\gamma_x)}$, then $d_T$ is the exponent of the temperature $T$ in the modified Stefan-Boltzmann law: 
\begin{equation}
\rho_{\gamma_{t}\gamma_{x}} \propto T^{1+3\frac{1+\gamma_t}{1+\gamma_x}}
\end{equation}
which can be obtained as usual deriving the logarithm of the thermodynamical partition function  ~\cite{termdim} with respect to the temperature $T$. Evidently, in our case we have that $\gamma_t = 0,\gamma_x = \frac{\gamma}{2}$ and, thus, we find $
d_T = 1 + \frac{6}{2+\gamma}$,
i.e. the thermal dimension agrees with the spectral dimension $d_T \equiv d_S$. \\
Finally, we can calculate also the Hausdorff dimension of momentum space that, if the duality with spacetime is not broken by quantum effects, should agree with both $d_S$ and $d_T$. As pointed out in Ref. ~\cite{momhaus}, a way to compute $d_H$ is to find a set of momenta that "linearize" the MDR. Given Eq. \eqref{mdr}, a possible choice is given by: 
\begin{equation}
k = \sqrt{p^2+\frac{2\alpha}{\gamma +2}l_{P}^{\gamma} p^{\gamma+2}}
\end{equation}  
In terms of these new variables $(E,k)$ the UV measure on momentum space becomes: 
\begin{equation}
\label{defmeas}
p^2 dpdE \longrightarrow k^{\frac{4-\gamma}{\gamma+2}} dk dE
\end{equation}
From the above equation \eqref{defmeas} we can read off $d_H$:
\begin{equation}
\label{hausdim}
d_H = 2 + \frac{4-\gamma}{\gamma+2} = 1 + \frac{6}{2+\gamma}
\end{equation}
that, evidently, coincides with both $d_S$ and $d_T$. Thus, no matter which definition of dimensionality is used, in the semi-classical limit (or in symmetry reduced models) of LQG the UV running is free of ambiguities since we have found that $d_S \equiv d_T \equiv d_H$. \\
The last consideration we want to make concerns what the number of UV dimensions can teach us about LQG. In the analysis we here reported, the value of $\gamma$ should be provided by the LQG corrections used to build $H^Q [N]$, which, though, are far from being unique. On the other hand, we mentioned that the number of dimensions runs to two in the UV, a prediction that seems to be model independent. Moreover, support in LQG has been also found by the studies of Refs.  ~\cite{lqg,calcoriti} under certain assumptions. It is evident from Eqs. \eqref{spectral}\eqref{hausdim} that in order to reproduce such a shared expectation we should take $\gamma = 4$. A recurring form of holonomy corrections both in spherically symmetric LQG ~\cite{hyppa,hyp2,holo1,splqg} as well as in LQC ~\cite{lqc,lqc1} is represented by the choice: $f(K) = \frac{\sin(\delta K)}{\delta}$ that implies $\beta = \cos(2\delta K)$.  In light of the above arguments, if we restrict to mesoscopic scales where the MDR is well approximated by the first-order correction, it is easy to realize that this implies $\gamma = 2$ (see Ref. ~\cite{kappalqg} for the explicit computation). In this way we would have $d = 2.5$ at scales near but below the Planck scale. We find rather encouraging the fact that we obtain a value which is greater than $2$, since such an outcome may signal that the descent from the classical value of 4 to the UV value is in progress but it has not been completed yet. In fact, at ultra-Planckian energies the Taylor expansion of the correction function $f(K)$ in series of powers of $\delta$ is no more reliable and, as a consequence, it can not fully capture the flowing of $d$. Therefore, we are led to conclude that the much-used polymerization of homogeneous connections, which is a direct consequence of evaluating the holonomies in the fundamental $j=\frac{1}{2}$ representation of $SU(2)$, realizes at least partially  the expected running of dimensions. Polymerizing configuration variables is used not only in LQC and in symmetry reduced contexts but also in the definition of the semi-classical regime of LQG, which is based on the introduction of spin-network states peaked around a single representation, in the majority of cases $j = \frac{1}{2}$. We have here shown that these quantum modifications can be related to the phenomenon of dimensional reduction. This link we have established gives us the possibility to constrain part of quantization ambiguities in LQG. In fact, the value of $d_{UV}$ fixes a specific choice of the parameter $\gamma$ in Eq. \eqref{beta}, thereby restricting the form of the allowed deformation functions $K \rightarrow f(K)$. In light of our analysis one should select only those modifications which are compatible with the prediction of a two-dimensional spacetime at ultra-short distances, i.e. those giving $2\lesssim d < 4$ (or equivalently $0 < \gamma \lesssim 4$) to a first approximation.

\section{OUTLOOK}

In the recent quantum-gravity literature there have been proposed basically three different definitions of dimension with the aim to generalize this notion for a quantum spacetime. It is well known that they all provide a possible characterization of the UV running but, in the majority of cases, they also give different outcomes for the value of the dimension. This clashes with the growing consensus on the fact that the phenomenon of dynamical dimensional reduction is a model independent feature of quantum gravity that gives the unique predictions $d_{UV} \simeq 2$. Thus, if both $d_S,d_T$ and $d_H$ are really proper definitions of the UV dimension, we would like to have that $d_S \equiv d_T \equiv d_H$.\\ In this paper we showed that this striking convergence can be accomplished in the case of the LQG approach. To achieve such a result, we relied on the deformations of the constraint algebra recently proposed in the framework of both effective spherically symmetric LQG and LQC. Remarkably, in the former case it has been proven that these quantum corrections leave traces in the Minkowski limit in terms of a DSR-like Poincaré algebra where the relevant deformations are functions of the spatial momentum. We here exploited these LQG-motivated deformations of the relativistic symmetries to infer the generic form of the MDR. Interestingly our MDR is qualitatively of the same type of two previously proposed MDRs for LQG  ~\cite{GambiniPullin,AlfaroTecotl}, namely the modifications affect only the momentum sector. From the MDR we derived both the spectral, the thermal and the Hausdorff dimensions proving that they all agree. Thus, in the top-down approach of LQG the desirable convergence of different characterization of the concept of dimensionality is accomplished. On the other had, the analysis we here reported may provide a guiding principle for the construction of bottom-up approaches. \\  Moreover, our analysis led us to give a contribution toward enforcing the fecund bond between theoretical formalisms and phenomenological predictions. In fact, we found that the simple polymerization of connections, which is much-used both in midi-superspace models and in LQC where semi-classical states are exploited to compute effective constraints, is able to generate the running of the dimension. In this way we have provided further evidence that the phenomenon of UV dimensional reduction can be realized also in the LQG approach, thereby confirming the results of previous studies. Remarkably, the value of $d_{UV}$ is sensible to the specific choice of quantum corrections which are considered in the model. Therefore, we pointed out that there is an observable whose value can be used to select a particular form for the quantum correction functions, thereby reducing the LQG quantization ambiguities. In particular, we showed that the evaluation of the Hamiltonian constraint over semi-classical states peaked at $j=\frac{1}{2}$ (that can be also implemented with the substitution $K \rightarrow \frac{\sin(\delta K)}{\delta}$ at an effective level) corresponds to $d_{UV} = 2.5$. Since we inferred such a value from the parametrization of the first LQG correction \eqref{beta} (which corresponds to a second order correction in the Planck length $\thicksim l_p^2$), then it is reasonable to regard our result as a first approximation that can not capture the ending outcome of the dimensional running. From this perspective we observed that obtaining a dimension grater than $2$ at energies near but below the Planck scale might be significant, because it can be read as a hint that the dimension is flowing to the "magic" number of $2$ that we can expect to be reached in the deep UV. Developing the off-shell constraint algebra in full generality without any symmetrical reduction or semi-classical approximation would be of fundamental importance to extend our observations to the full LQG framework.

\section*{Acknowledgments.} The author is grateful to Giovanni Amelino-Camelia and Giorgio Immirzi for useful and critical discussions during the first stages of the elaboration of this work. He also wants to thank Gianluca Calcagni for carefully reading the manuscript and giving many helpful comments that improved this work. Finally, he acknowledges referees for their constructive suggestions.

\end{document}